\documentclass{aa}
\usepackage{graphicx}
\usepackage[varg]{txfonts}
\usepackage[colorlinks = true,
            linkcolor = blue,
            urlcolor  = blue,
            citecolor = blue]{hyperref}
\usepackage{euclid}

\newcommand{\DateOrder}[1]{}

\usepackage[nameinlink,noabbrev]{cleveref}
\Crefname{eqnarray}{Eq.}{Eqs.}
\Crefname{section}{Sect.}{Sects.}
\Crefname{figure}{Fig.}{Figs.}
\crefname{equation}{Equation}{Equations}
\crefname{section}{Section}{Sections}
\crefname{figure}{Figure}{Figures}

\titlerunning{Voids on cosmology tensions}

\begin{document}

\title{The perspective of voids on rising cosmology tensions}

\newcommand{\orcid}[1]{} 
\author{S.~Contarini\orcid{0000-0002-9843-723X}$^{1,2,3}$\thanks{\email{sofia.contarini3@unibo.it}}, A.~Pisani\orcid{0000-0002-6146-4437}$^{4,5,6}$, N.~Hamaus\orcid{0000-0002-0876-2101}$^{7}$, F.~Marulli\orcid{0000-0002-8850-0303}$^{1,2,3}$, L.~Moscardini\orcid{0000-0002-3473-6716}$^{1,2,3}$, M.~Baldi\orcid{0000-0003-4145-1943}$^{1,2,3}$}

\institute{$^{1}$ Dipartimento di Fisica e Astronomia "Augusto Righi" - Alma Mater Studiorum Universit\`{a} di Bologna, via Piero Gobetti 93/2, I-40129 Bologna, Italy\\
$^{2}$ INFN-Sezione di Bologna, Viale Berti Pichat 6/2, I-40127 Bologna, Italy\\
$^{3}$ INAF-Osservatorio di Astrofisica e Scienza dello Spazio di Bologna, Via Piero Gobetti 93/3, I-40129 Bologna, Italy\\
$^{4}$ Center for Computational Astrophysics, Flatiron Institute, 162 5th Avenue, 10010, New York, NY, USA\\
$^{5}$ The Cooper Union for the Advancement of Science and Art, 41 Cooper Square, New York, NY 10003, USA\\
$^{6}$ Department of Astrophysical Sciences, Peyton Hall, Princeton University, 4 Ivy Lane, Princeton, NJ 08544, USA\\
$^{7}$ Universit\"ats-Sternwarte M\"unchen, Fakult\"at f\"ur Physik, Ludwig-Maximilians-Universit\"at, Scheinerstrasse 1, 81679 M\"unchen, Germany}

\date{\vspace{-5ex}}

\authorrunning{S. Contarini et al.} 

\abstract{We investigate the main tensions within the current standard model of cosmology from the perspective of the main statistics of cosmic voids, using the final BOSS DR12 data set. For this purpose, we present the first estimate of the $S_8\equiv \sigma_8\sqrt{\Omega_{\rm m}/0.3}$ and $H_0$ parameters obtained from void number counts and shape distortions. To analyze void counts we relied on an extension of the popular volume-conserving model for the void size function, tailored to the application on data, including geometric and dynamic distortions. We calibrated the two nuisance parameters of this model with the official BOSS collaboration mock catalogs and propagated their uncertainty through the statistical analysis of the BOSS void number counts. The constraints from void shapes come from the study of the geometric distortions of the stacked void-galaxy cross-correlation function. In this work we focus our analysis on the $\Omega_{\rm m}$--$\sigma_8$ and $\Omega_{\rm m}$--$H_0$ parameter planes and derive the marginalized constraints $S_8 = 0.813^{+0.093}_{-0.068}$ and $H_0 = 67.3^{+10.0}_{-9.1} \ \mathrm{km} \ \mathrm{s}^{-1} \ \mathrm{Mpc}^{-1}$, which are fully compatible with constraints from the literature. These results are expected to notably improve in precision when analyzed jointly with independent probes and will open a new viewing angle on the rising cosmological tensions in the near future. }

\maketitle

\section{Introduction}
In recent years anomalies in the observations of the Universe at large scales have puzzled the scientific community. These are expressed as statistically relevant discrepancies, or tensions, between the cosmological constraints obtained from late (i.e., low-redshift) and early (i.e., high-redshift) probes.

The most significant tension affects the estimate of the Hubble constant $H_0$ describing the present expansion rate of the Universe. In particular, the   Hubble tension concerns the mismatch between the Planck collaboration results obtained from the cosmic microwave background (CMB) anisotropies \citep{Planck2018_results} and the direct local distance ladder measurements based on type Ia supernovae \citep[SN Ia,][]{Riess2022}. These two estimates are in disagreement by about $5\sigma$ \citep[see, e.g.,][]{DiValentino2021a}.
Second in order of importance is the tension on the matter clustering strength, often quantified by the parameter ${S_8 \equiv \sigma_8\sqrt{\Omega_{\rm m}/0.3}}$, derived from the matter density parameter $\Omega_{\rm m}$ and the root mean square of density fluctuations 
on an $8 \ h^{-1}{\rm Mpc}$ comoving scale $\sigma_8$. In this case the discrepancy of the Planck satellite measurements relates to low-redshift probes, such as weak gravitational lensing and galaxy clustering, with a statistical disagreement at the level of $2$-$3\sigma$ \citep[see, e.g.,][]{DiValentino2021b}.

These observational tensions may reveal the presence of systematic errors in the cosmological measurements or, alternatively, the requirement to modify the current benchmark Lambda-cold dark matter ($\Lambda$CDM) model of cosmology. This has led to the proposal of a plethora of alternatives to the $\Lambda$CDM model, involving physics beyond the standard model of particle physics \citep[see][and references therein]{CI2022}.
Although some of the proposed models alleviate one or more discrepancies \citep[see, e.g.,][]{Panday2020,DiValentino2021solutions,Schoneberg2022}, we are still longing for a definitive solution that provides a comprehensive and satisfactory description of all the cosmological observables.

In this context the darkest regions of our Universe may supply a fundamental and independent contribution,  shedding light on the rising cosmological tensions. These vast underdense zones, scarcely populated by galaxies and other luminous objects, are commonly referred to as cosmic voids. Within the last decade, voids have begun to assert their relevance as cosmological probes \citep{pisani2019,Moresco2022}. Recent works have indeed exploited voids to test the standard cosmological model \citep[e.g.,][]{hamaus2016,hamaus2020,aubert2020,nadathur2020,woodfinden2022,kovacs2022}, and others have provided promising forecasts on the constraining power expected from different void statistics with the upcoming redshift surveys, such as \textit{Euclid} \citep{Laureijs2011, Amendola2018,EuclidHamaus2021,Contarini2022A,Bonici2022}.

Nevertheless, to date the void size function as a first-order statistic of voids has never been exploited to derive cosmological constraints. In this work we extend the results presented in \cite{Contarini2022BOSS}, where we performed a statistical analysis of the size distribution of voids identified in the final data release (DR12) of the Baryon Oscillation Spectroscopic Survey (BOSS, \citealp{SDSS2011,BOSSDR12}). We focus on the main cosmological tensions, and we add to our results the contribution from the constraints achieved by \cite{hamaus2020}, who relied on that same data to analyze void shape distortions. This combination  provides us with a first estimate of the $S_8$ and $H_0$ parameters derived from cosmic voids. We discuss our results in the current context of cosmology, featuring the constraints from different surveys and cosmological probes.

This paper is structured as follows. In \Cref{sec:Theory_and_data} we present the theory of the void size function, together with the galaxy and void catalogs we employed. In \Cref{sec:analysis} we describe the cosmological analysis performed in this work. The constraints obtained in this context are then presented in \Cref{sec:Results} and compared with other important results from the literature. Finally, in \Cref{sec:Conclusions} we draw our conclusions of this work.

\section{Theory and data sets} \label{sec:Theory_and_data}
\subsection{Void size function model}\label{subsec:theory}
The void size function model predicts the comoving number density of voids as a function of their size. It was originally developed by \cite{SVdW2004} on the basis of the excursion-set theory and at its core relies on the multiplicity function
\begin{equation}\label{eq:multiplicity}
\begin{gathered}
f_{\ln \sigma_{\rm R}} = 2 \sum_{j=1}^{\infty} \, \exp{\bigg(-\frac{(j \pi x)^2}{2}\bigg)} \, j \pi x^2 \, \sin{\left( j \pi \mathcal{D} \right)}\, ,\\
\text{with} \quad \mathcal{D} = \frac{|\delta_\mathrm{v}^\mathrm{L}|}{\delta_\mathrm{c}^\mathrm{L} + |\delta_\mathrm{v}^\mathrm{L}|}\, \quad 
\text{and} \quad x = \frac{\mathcal{D}}{|\delta_\mathrm{v}^\mathrm{L}|} \sigma_{\rm R} \, ,
\end{gathered}
\end{equation}
which describes the volume fraction of the Universe occupied by cosmic voids. In \Cref{eq:multiplicity} all the quantities are expressed in linear theory, as indicated by the superscript ``L''. In particular,
$\sigma_{\rm R}$ is the root mean square variance of linear matter perturbations on a scale $R_\mathrm{L}$, while $\delta_\mathrm{c}^\mathrm{L}$ and $\delta_\mathrm{v}^\mathrm{L}$ are the density contrasts required for the formation of dark matter halos and cosmic voids, respectively.
Only the latter threshold significantly affects the predicted void size distribution at large scales and determines the linear density contrast embedded inside cosmic voids. We note the importance of this quantity  below, and refer  to \cite{Contarini2022BOSS} for a more complete description.

To calculate the density of cosmic voids as a function of their radii $R$ in the nonlinear regime, \cite{jennings2013} proposed the following expression:
\begin{equation}
\frac{\diff n}{\diff \ln R} = \frac{f_{\ln \sigma_{\rm R}}}{V(R)} \, \frac{\diff \ln \sigma_{\rm R}^{-1}}{\diff \ln R_\mathrm{L}} \biggr \rvert_{R_\mathrm{L} = R_\mathrm{L}(R)}\, ,
\end{equation}
which is called the  volume-conserving model (hereafter Vdn). As the name suggests, it relies on the conservation of the total volume $V$, occupied by voids in the transition from the linear to the nonlinear regime. The Vdn model has been tested successively in different works \citep[see, e.g.,][]{jennings2013,ronconi2019,verza2019,Contarini2020}, but for its application to voids identified in a biased distribution of tracers (e.g., galaxies and clusters of galaxies) a modification of its main assumptions is required. 

To take into account the change in the density contrast when voids are traced by biased objects, \cite{contarini2019} introduced a simple parameterization of the underdensity threshold of the Vdn model. This strategy was then revisited in \cite{Contarini2022BOSS} to take into account the degeneracy of the tracer effective bias with the normalization of the matter density fluctuations $\sigma_8$, leading to the following parameterization of the void density threshold:
\begin{equation}\label{eq:thr_conversion}
\delta_\mathrm{v,DM}^\mathrm{NL} = \frac{\delta_\mathrm{v,tr}^\mathrm{NL}}{\mathcal{F}(b_\mathrm{eff}, \sigma_8)} \,,
\end{equation}
with
\begin{equation}\label{eq:Fbeff}
\mathcal{F}(b_\mathrm{eff} \, \sigma_8) = C_\mathrm{slope} \, b_\mathrm{eff} \, \sigma_8 + C_\mathrm{offset} \, .
\end{equation}
In this equation we use the superscript ``NL'' to highlight quantities computed in nonlinear theory;  $\delta_\mathrm{v,tr}^\mathrm{NL}$ is the density contrast used to define voids in the tracer density field (tr), while $\delta_\mathrm{v,DM}^\mathrm{NL}$ is the corresponding value in the matter density field (DM, i.e., dark matter particles); the function $\mathcal{F}(b_\mathrm{eff} \, \sigma_8)$ parameterizes the action of the tracer effective bias $b_\mathrm{eff}$, which depends on redshift and on the selected tracers (e.g., their host-halo mass); and $C_{\rm slope}$ and $C_{\rm offset}$ are redshift-independent coefficients of the linear function $\mathcal{F}$ and their dependence on the cosmological model can be considered negligible \citep{Contarini2020}.
Moreover, \cite{Contarini2022A} demonstrated that the parameterization presented in \Cref{eq:thr_conversion,eq:Fbeff} is effective in encapsulating redshift-space distortions on cosmic voids (i.e., the enlargement of their observed radii) caused by the peculiar motions of their tracers.

The presented extension of the Vdn model predicts the size function of voids identified by biased tracers in redshift space with good accuracy, provided the calibration of the nuisance parameters of the model, $C_{\rm slope}$ and $C_{\rm offset}$, is performed via mock catalogs designed to reproduce the target tracer population \citep[see][for further details]{Contarini2020}. Hereafter, we   refer to this theoretical framework as the extended Vdn model.

\subsection{Galaxy and void catalogs}\label{subsec:datsets}
In this work we analyze the BOSS DR12 data set \citep{Reid2016}, composed of the two target selections LOWZ and CMASS featuring more than one million galaxies with spectroscopic redshifts. We also use 100 realizations of the MultiDark PATCHY mocks \citep{kitaura2014,BOSSDR12_clustering,MultiDark2016,RT2016}, specifically designed to mimic the properties of the BOSS galaxies, to calibrate the nuisance parameters of the extended Vdn model (see \Cref{subsec:theory}). The effectiveness of these mocks in reproducing the main clustering properties of different populations of objects is demonstrated by \cite{Kitaura2016} and \cite{RT2016}, who verify the accuracy of the power spectrum, two- and three-point correlation functions down to scales of tens of megaparsec.

Following the procedure of \cite{Contarini2022BOSS}, we divided the catalogs into two redshift bins ($0.2 < z \le 0.45$ and $0.45<z<0.65$), and measured $b_\mathrm{eff} \, \sigma_8$ for each case, as required in \Cref{eq:Fbeff}. We estimated this quantity by modeling the multipoles of the galaxy two-point correlation function with the prescriptions of \cite{Taruya2010} (see Appendix A of \citealp{Contarini2022BOSS} for further details). 
For the BOSS galaxies this yielded $b_\mathrm{eff} \, \sigma_8 = 1.36 \pm 0.05$ in the interval $0.2 < z \le 0.45$ and $b_\mathrm{eff} \, \sigma_8 = 1.28 \pm 0.06$ in the $0.45 < z < 0.65$ interval. 

To identify cosmic voids in the distribution of real and mock galaxies we applied the public Void IDentification and Examination toolkit\footnote{\url{https://bitbucket.org/cosmicvoids/vide_public}} \citep[\texttt{VIDE},][]{vide}, based on the code ZOnes Bordering On Voidness \citep[\texttt{ZOBOV},][]{zobov}. \texttt{VIDE} exploits the Voronoi tessellation technique to approximate a continuous density field and identifies its local minima. From this a catalog of voids was built by means of a watershed algorithm \citep{platen2007}. We ran \texttt{VIDE} on the entire galaxy distribution from the northern and southern Galactic hemispheres and assumed the same cosmological model used to build the MultiDark PATCHY simulations to convert redshifts to comoving distances (i.e., a Planck2013 cosmology; \citealt[][see also \Cref{sec:analysis}]{Planck2013}).

Once voids were identified, we processed the extracted sample to match the theoretical definition used in the void size function model. In particular, we applied the cleaning procedure developed by \cite{ronconi2017}, publicly available in the \textit{free software} C++/Python libraries \texttt{CosmoBolognaLib}\footnote{\url{https://gitlab.com/federicomarulli/CosmoBolognaLib}}\citep{CBL}, which rescales voids such that they exhibit a fixed density contrast of $\delta_\mathrm{v,tr}^\mathrm{NL}=-0.7$. This choice of underdensity threshold is not unique, but is especially suited to obtain a statistically reliable void sample \citep[see, e.g.,][for further details]{Contarini2022A}. The value of $\delta_\mathrm{v,tr}^\mathrm{NL}$ is the only free parameter in the cleaning algorithm, and it must be chosen in agreement with the underdensity threshold used in the void size function model, presented in \cref{eq:thr_conversion}. Other internal values can be set to speed up the computation, but do not have significant effects on the output. Therefore, we consider the impact of parameter settings during the finding and cleaning procedures to be negligible \citep[also following  the approach of, e.g.,][]{hamaus2020,EuclidHamaus2021,Contarini2022A,Contarini2022BOSS}.

Nevertheless, the minimum radius of the void catalog must be chosen accurately to avoid selecting the spatial scales affected by void count incompleteness. This is usually done by discarding all the voids that, after the cleaning procedure, exhibit a radius smaller than a factor $f_\mathrm{cut}$ times the mean separation of the tracers $n_\mathrm{tr}(z)^{-1/3}$ \citep{jennings2013,verza2019,contarini2019,Contarini2020,Contarini2022A,Pelliciari}. We adopted the conservative choice of $f_\mathrm{cut}=2.5$, prioritizing the robustness of the result over the strength of the derived cosmological constraints. We refer   to Appendix C.3 of \cite{Contarini2022BOSS} for further details about the impact of the minimum void radius selection.

We then measured the  number counts of the cleaned voids as a function of their size.
We first used the average PATCHY void counts to calibrate the values of $C_{\rm slope}$ and $C_{\rm offset}$ through a Markov chain Monte Carlo (MCMC) analysis. The best-fit values we obtained for the coefficients of $\mathcal{F}(b_\mathrm{eff} \, \sigma_8)$ are $C_\mathrm{slope}=-1.62^{+0.66}_{-0.73}$ and $C_\mathrm{offset}=4.30^{+0.94}_{-0.86}$, but we refer   to \cite{Contarini2022BOSS} for a complete presentation of these results.
We then exploited the calibration performed with the PATCHY mocks to model the BOSS void number counts, sampling the posterior distribution of the free parameters of the extended Vdn model (see \Cref{sec:analysis}). We note that the effectiveness of the analysis we apply in this work is tested in Appendix B of \cite{Contarini2022BOSS}, where the methodology is thoroughly validated on PATCHY mocks. Additionally, in Appendix C.1 of the same paper, we explore the impact of the priors applied to the nuisance parameters $C_{\rm slope}$ and $C_{\rm offset}$, demonstrating their nearly negligible effect in constraining $\Omega_\mathrm{m}$.

\section{Analysis}\label{sec:analysis}

The cosmological analysis we carry out in this work is based on Bayesian inference. We take as data set the void counts extracted from the BOSS DR12 and assume a Poissonian likelihood \citep{sahlen2016}. Our model is based on the extended Vdn theory, so it depends on the cosmological model and on the nuisance parameters $C_{\rm slope}$ and $C_{\rm offset}$ (see \Cref{subsec:theory}).

We consider a flat $\Lambda$CDM model with fiducial parameters given by the  Planck2013 results \citep{Planck2013} to be consistent with those used to build the MultiDark PATCHY simulations: $\Omega_{\rm m}=0.307115$, $H_0=67.77 \ \mathrm{km} \ \mathrm{s}^{-1} \ \mathrm{Mpc}^{-1}$, $\Omega_{\rm b}=0.048206$, $n_{\rm s}=0.9611$, and $\sigma_8=0.8288$. Our statistical analysis is aimed at investigating the degeneracies in the $\Omega_{\rm m}$--$\sigma_8$ and $\Omega_{\rm m}$--$H_0$ planes, so we focus on two sets of cosmological parameters, the first characterized by wide uniform priors for $\Omega_{\rm m}$ and $\sigma_8$, and the second for $\Omega_{\rm m}$ and $H_0$. In both cases we marginalize over the remaining cosmological parameters by assigning to them uniform priors centered on our fiducial cosmology, with a total width of ten times the $68\%$ uncertainties provided by the  Planck2018 constraints \citep{Planck2018_results}. 
We also marginalize over $C_{\rm slope}$ and $C_{\rm offset}$ by considering their joint posterior distribution obtained from the calibration with PATCHY mocks.

\begin{figure*}
\centering
\includegraphics[height=8.cm]{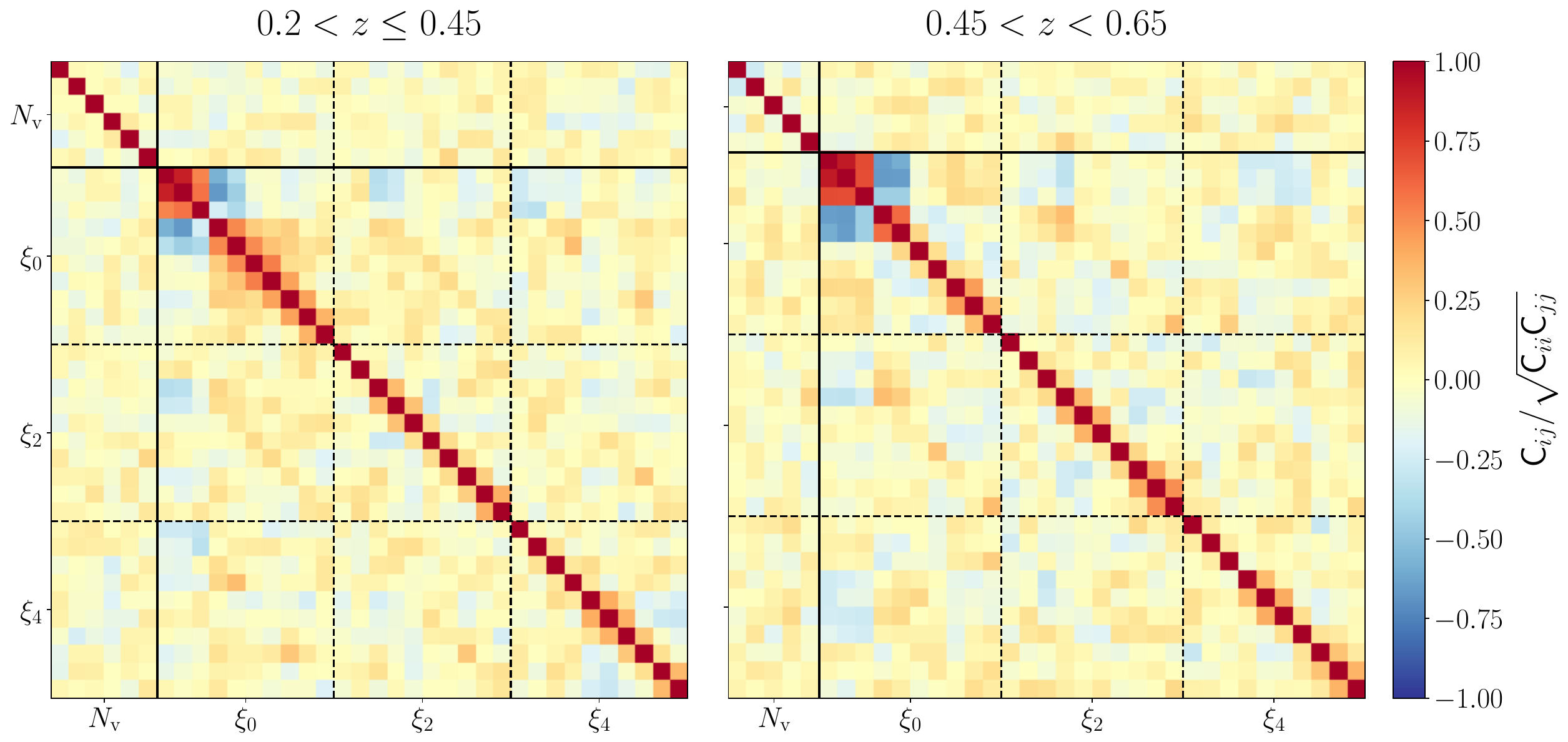}
\caption{Covariance matrix of void counts and stacked void-galaxy cross-correlation function (monopole, $\xi_0$, quadrupole, $\xi_2$, and hexadecapole, $\xi_4$), computed via $100$ PATCHY mocks. The matrix is normalized by its diagonal components and the two panels correspond to the redshift bins used in this analysis. The solid black lines divide the void size function data vector with the void-galaxy cross-correlation function ones, while the dashed lines separate the different multipoles from each other.}
\label{fig_cov}
\end{figure*}

We finally sample the posterior distribution of the considered parameter sets following an MCMC technique. At each step of the MCMC we consider a new set of cosmological parameters, compute the void size function, and rescale the predicted void radii to model the effect of geometric distortions acting at the mean redshift of the selected void sample.
These distortions arise when assuming an incorrect cosmology in the redshift-distance conversion, causing a distortion of the observed void shape along the directions parallel and perpendicular to the line of sight. The resulting anisotropy is referred to as the  Alcock-Paczyński (AP) effect \citep{AP1979} and can be modeled via two scaling parameters, $q_\parallel$ and $q_\perp$ \citep[e.g.,][]{Sanchez2017b}.
We indicate with $r^*_\parallel$ and $r^*_\perp$ the observed comoving distances between two objects at redshift $z$, projected along the parallel and perpendicular line-of-sight directions, respectively. To obtain their corresponding length in the true cosmological model, we use
\begin{equation}
r_\parallel = \frac{H^*(z)}{H(z)} \, r^*_\parallel \equiv q_\parallel \, r^*_\parallel \,, \quad
r_\perp = \frac{D_{\rm A}(z)}{D^*_{\rm A}(z)} \, r^*_\perp \equiv q_\perp \, r^*_\perp \, .
\end{equation}
In these equations the starred quantities are computed assuming the fiducial cosmology: $H(z)$ is the Hubble parameter and $D_{\rm A}(z)$ the comoving angular-diameter distance, which in flat $\Lambda$CDM are defined as
\begin{equation}
H(z)=H_0\left[ \Omega_{\rm m}(1+z)^3 +1-\Omega_{\rm m}\right]^{1/2}\,
\end{equation}
and
\begin{equation}
D_{\rm A}(z)=\int^z_0 \frac{c}{H(z')}{\rm d}z' \,,
\end{equation}
respectively. The inclusion of geometric distortions in the void size function model causes a shift in the predicted radius distribution that depends on the discrepancy between the true and the assumed fiducial cosmology.
We finally note  that we did not account for super-sample covariance in our model, which has recently   been demonstrated to only marginally affect the void size function \citep{Bayer2022}.

To gather additional cosmological information from cosmic voids, we present in \Cref{subsec:2Dconstraints} the constraints derived by adding to our analysis the contribution of the void-galaxy cross-correlation function measured in the same BOSS DR12 data set. Specifically, we rely on the findings of \cite{hamaus2020}: utilizing voids as standard spheres to measure geometric distortions, these authors constrain  the growth rate of structure and provide  a precise estimate of the total matter density parameter (i.e., $\Omega_\mathrm{m} = 0.312 \pm 0.020)$. We exploit this constraint by imposing a Gaussian prior on $\Omega_\mathrm{m}$ in the MCMC analysis. This approach is equivalent to multiplying the resulting posterior distributions from the two analyses, assuming these void statistics to be completely independent. This assumption is supported by the results we present in the following section.

\section{Results}\label{sec:Results}

\subsection{Covariance with void-galaxy cross-correlation function}\label{subsec:covariance}

In this section we present the covariance matrix of void counts $N_\mathrm{v}$ and void-galaxy cross-correlation function multipoles $\xi_\ell$, with the aim of evaluating the cross-correlation between the two observables.
A similar analysis is performed by \cite{Kreisch2021}, and demonstrates the low cross-correlation between void size function and the stacked void-halo cross-correlation function in real space. We note  however that, conversely to what is presented in \cite{Kreisch2021}, in this work the sample of voids employed for the study of number counts has undergone a cleaning procedure (see \Cref{subsec:datsets}). This  is expected to further reduce the correlation between the two void statistics since the samples of voids considered are characterized by different features.

For this analysis we use  the void catalogs extracted with \texttt{VIDE} from MultiDark PATCHY mocks (see \Cref{subsec:datsets}) and compute  the void counts in the redshift bins $0.2 < z \le 0.45$ and $0.45<z<0.65$. These counts are the same as those presented in \cite{Contarini2022BOSS} and were derived from a cleaned sample of voids. In the same redshift bins, we measure the stacked void-galaxy cross-correlation function from the raw void catalogs, using the prescriptions adopted in \cite{hamaus2020}. 
With respect to the latter, we lowered the number of void-galaxy separation bins to ten in order to reduce the statistical noise. However, we verified the stability of the results by also computing the void-galaxy cross-correlation multipoles with $20$ bins.

Our vector of observables $\Vec{O}$ is therefore given by
\begin{equation}
    \Vec{O} = \{N_\mathrm{v}(R_1), ... , N_\mathrm{v}(R_n), \xi_\ell(s_1), ... , \xi_\ell(s_m)\} \, ,
\end{equation}
where $\xi_\ell$ is composed of the three even multipoles $\ell = (0, 2, 4)$, and $n, m$ correspond to the index of the largest void radius and void-galaxy separation bins. 
We estimate the covariance matrix using $N_\mathrm{cov}=100$ mock realizations as
\begin{equation}
    C_{\alpha\beta} = \langle (O_\alpha-\langle O_\alpha\rangle ) \, (O_\beta-\langle O_\beta\rangle )\rangle \, ,
\end{equation}
where $\alpha$ and $\beta$ represent our two void observables.

In \Cref{fig_cov} we present the correlation matrix $\mathrm{Corr}(O_\alpha, O_\beta) \equiv C_{\alpha\beta}/\sqrt{C_{\alpha\alpha}C_{\beta\beta}}$ obtained from the PATCHY mocks. This is shown with the purpose of investigating possible correlation between the void statistics $N_\mathrm{v}$ and $\xi_\ell$. We note that the cross-terms $N_\mathrm{v}-\xi_0$, $N_\mathrm{v}-\xi_2$, and $N_\mathrm{v}-\xi_4$ are close to zero, even though they are  affected by small statistical fluctuations. We verified that when applying a tapering factor \citep{PazSanchez2015} of around $50$, these terms are perfectly consistent with zero.

We conclude that our measures of void counts and void-galaxy cross-correlation function multipoles do not exhibit a statistically relevant covariance. For this reason, in the following analysis we consider these observables to be  fully independent.

\begin{figure*}
\centering
\includegraphics[height=6.5cm]{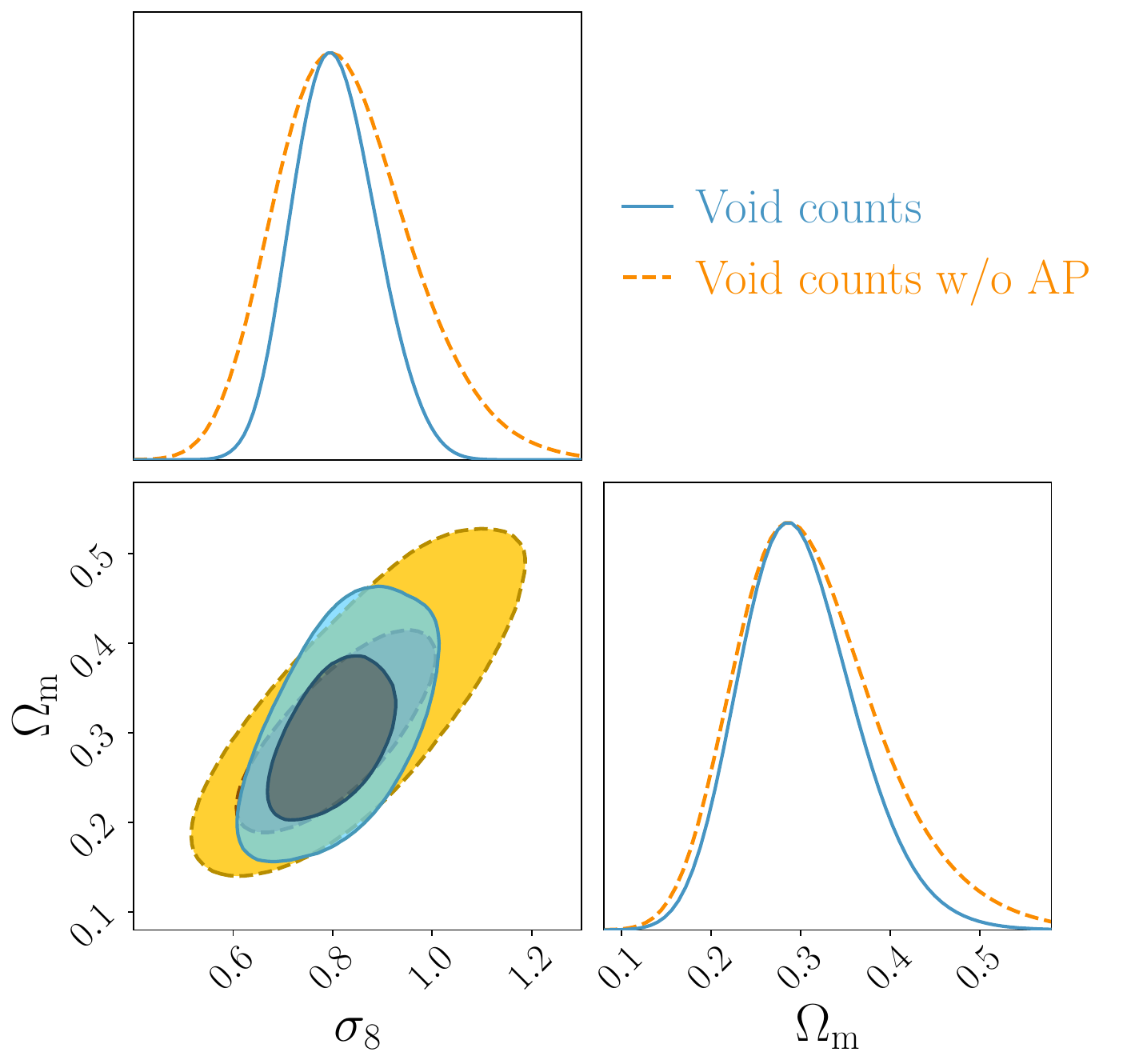}\hspace{0.1cm}
\hspace{1cm}
\includegraphics[height=6.5cm]{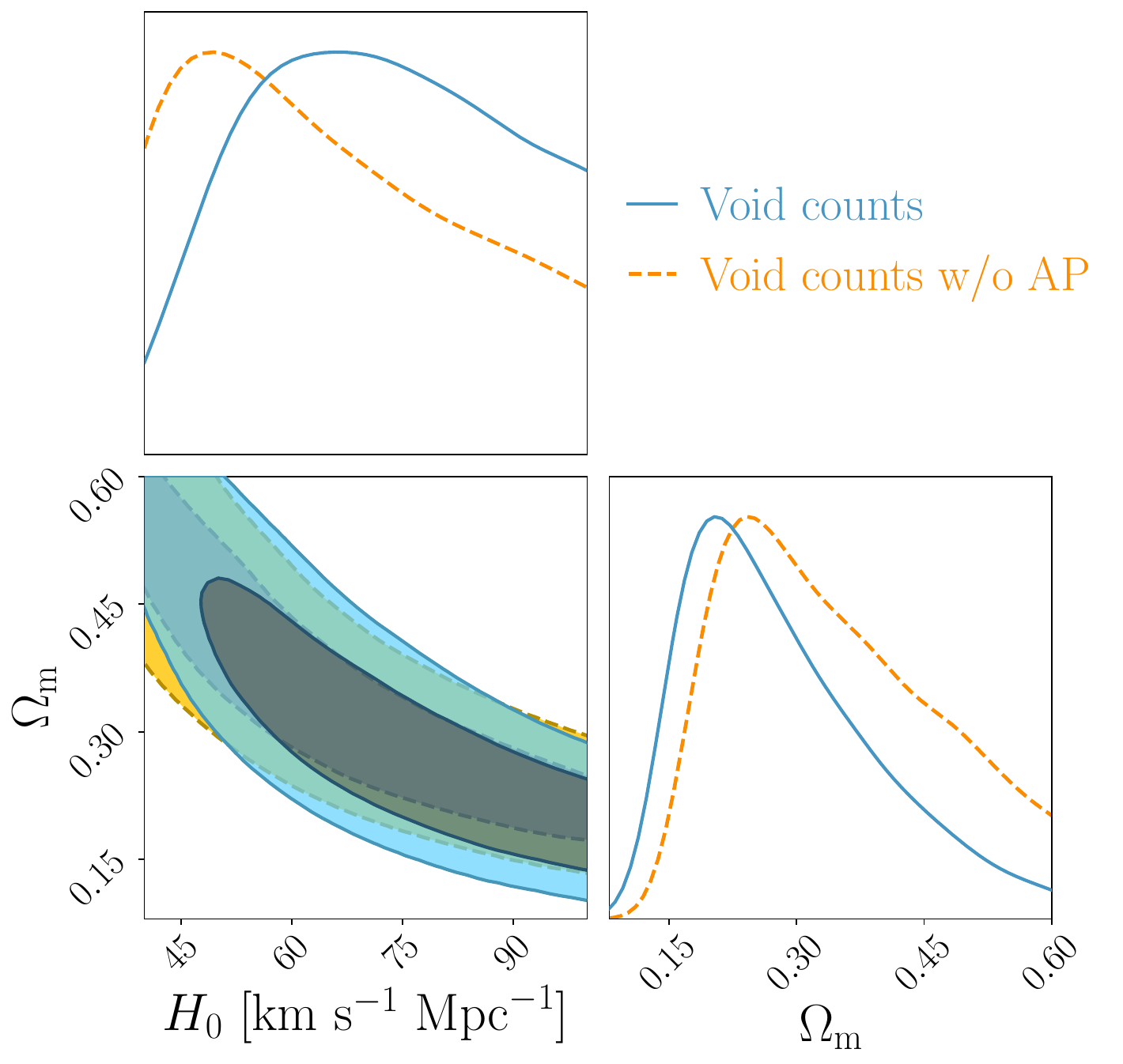}
\caption{$68\%$ and $95\%$ confidence contours from void counts (in light blue) compared to the confidence contours obtained without including the void radius correction for the AP effect (in orange, see \Cref{sec:analysis}). Shown on the left are the cosmological constraints for the $\Omega_{\rm m}$--$\sigma_8$ parameter plane, and on the right those for the $\Omega_{\rm m}$--$H_0$ parameter plane. \label{figAP}}
\end{figure*}

\subsection{Cosmological constraints from cosmic voids}\label{subsec:2Dconstraints}

In this section we present the main results of the analysis, expressed as new cosmological constraints from the BOSS DR12 data set. These results supplement the constraints presented in \cite{Contarini2022BOSS}.

In \Cref{figAP} we present the cosmological constraints from the void size function, in both the parameter planes analyzed, namely $\Omega_{\rm m}$--$\sigma_8$ and $\Omega_{\rm m}$--$H_0$. As a reference, we also report the constraints obtained when omitting the AP effect in our model (see \Cref{sec:analysis}). This allows us to quantify the impact of the geometric distortion correction on voids and the subsequent small improvement on the cosmological constraints. We note that this correction is generally negligible for collapsed and virialized structures, like galaxy clusters, because of their detachment from the Hubble expansion. We tested the effectiveness of the AP correction with PATCHY mocks by assuming different fiducial values of $\Omega_{\rm m}$, and we found an almost perfect match between the derived confidence contours (see Appendix C.2 of \citealp{Contarini2022BOSS}).

\begin{figure*}
\centering
\hspace{0.4cm}\includegraphics[height=6.5cm]{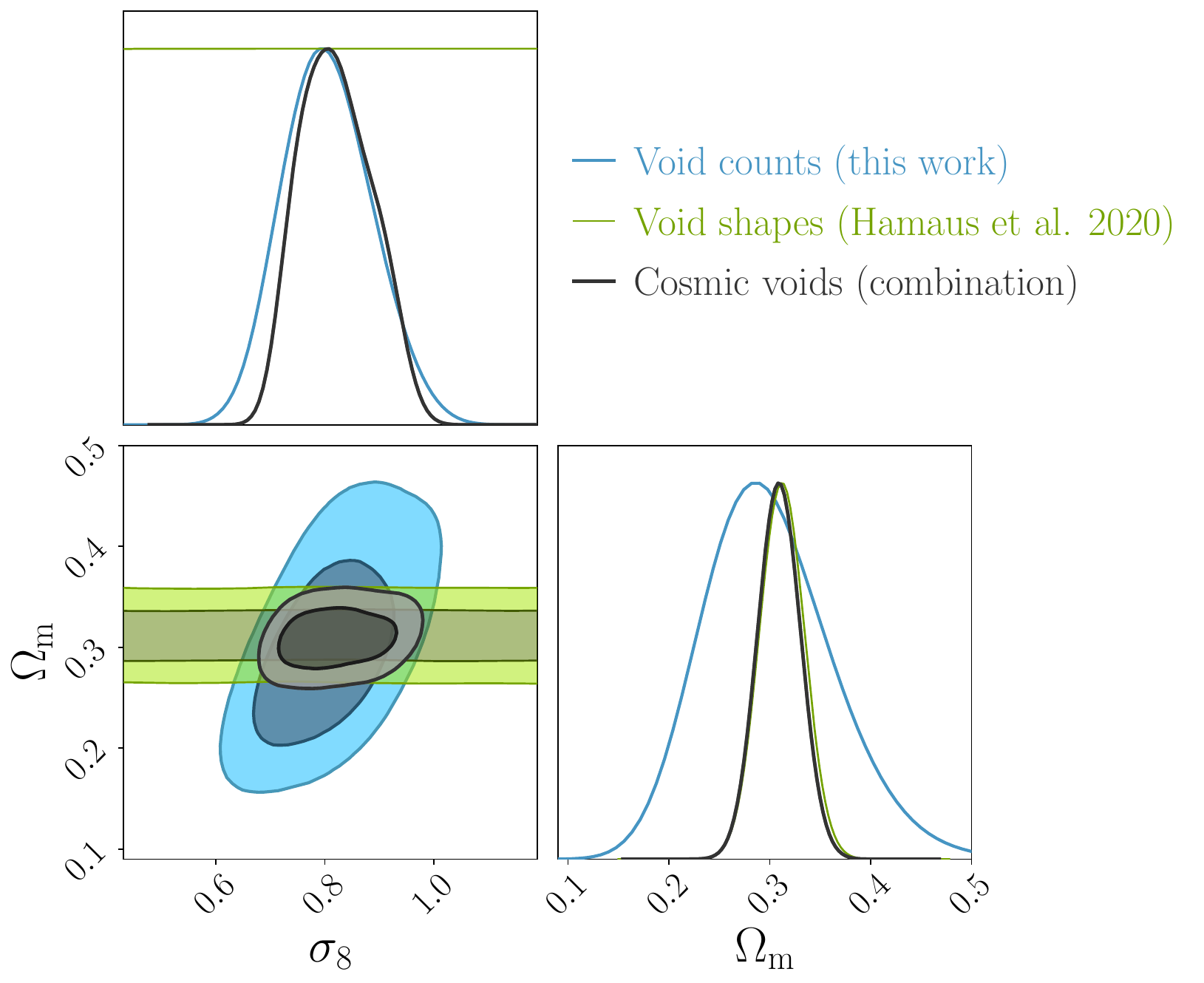}\hspace{0.1cm}
\hspace{0.3cm}
\includegraphics[height=6.5cm]{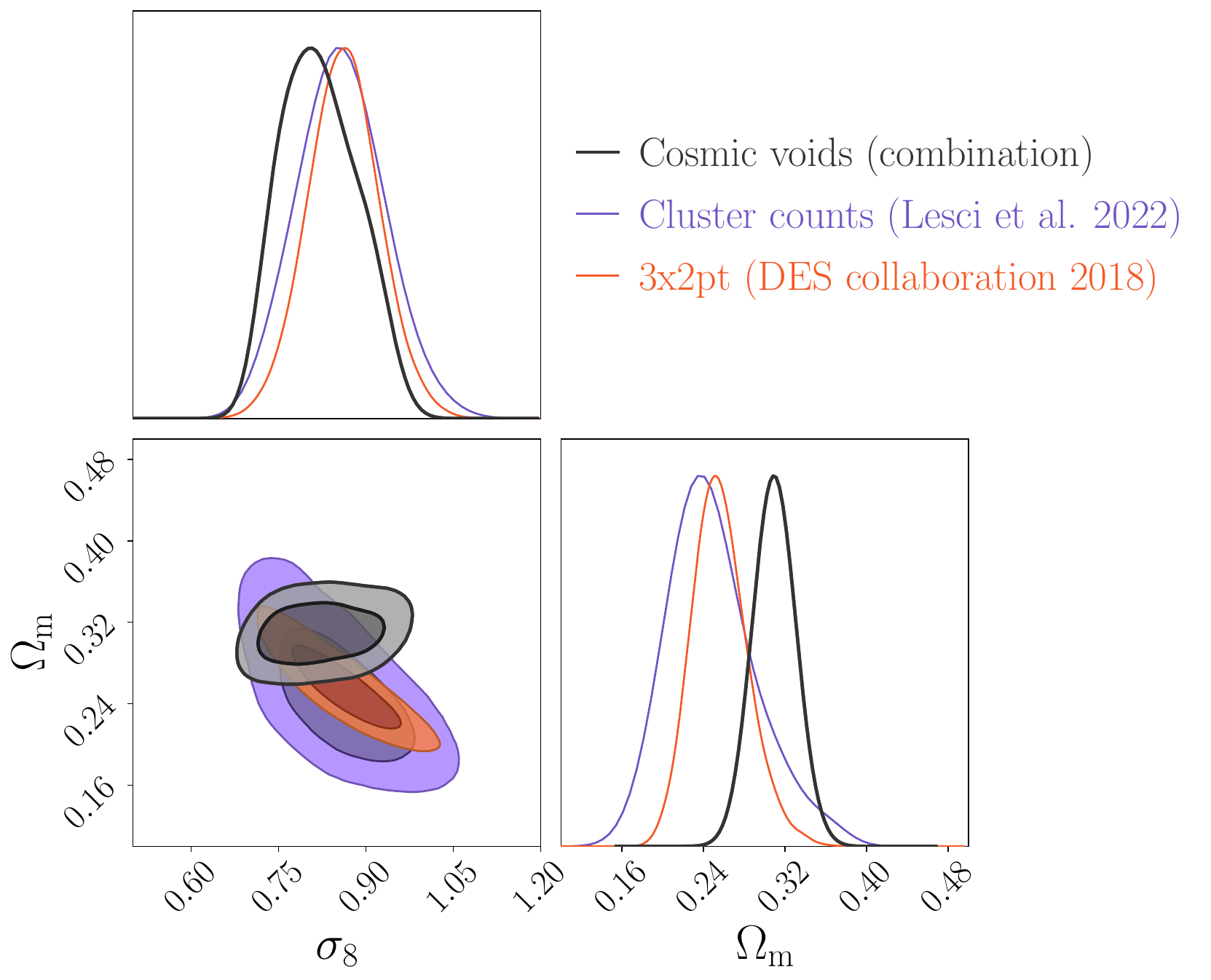}
\caption{$68\%$ and $95\%$ confidence contours on the $\Omega_{\rm m}$--$\sigma_8$ parameter plane and the corresponding projected posterior distributions.
Left: Confidence contours from void counts (light blue) and void shape distortions \citep[green, from][]{hamaus2020} of the BOSS DR12 voids, and their combination (black) as independent constraints. Right: Comparison of the results from cosmic voids with other cosmological probes, cluster counts \citep[purple, from][]{Lesci2022a} and the DES Y1 3x2pt analysis \citep[red, from][]{DES2018_3x2pt}.}\label{figS8}
\end{figure*}

We now compare our results with those obtained with other void statistics and cosmological probes, starting by focusing on the $\Omega_{\rm m}$--$\sigma_8$ parameter plane. In the left panels of \Cref{figS8} we report our constraints from void counts together with the estimate of $\Omega_{\rm m}$ from \cite{hamaus2020} from void shape distortions, already introduced in \Cref{sec:analysis} and reported here as a continuous band. In the same plot we present the product of the posterior density probability of the two void statistics. With the contribution from void shape distortions we obtain $\Omega_{\rm m} = 0.308^{+0.021}_{-0.018}$ and $\sigma_8 = 0.809^{+0.072}_{-0.068}$.
In the right panels we compare this  result with the confidence contours computed by \cite{Lesci2022a} using the mass function of clusters identified in the third data release of the Kilo Degree Survey (KiDS-DR3) and those of the combined analysis of galaxy clustering and weak gravitational
lensing\footnote{This analysis deals with the modeling of three two-point functions: the cosmic shear correlation function, the galaxy angular auto-correlation, and the galaxy-shear cross-correlation, and is commonly abbreviated as 3x2pt.} performed by the \cite{DES2018_3x2pt} using data from the first year of observations (DES Y1).

\begin{figure*}
\centering
\hspace{0.7cm}\includegraphics[height=6.5cm]{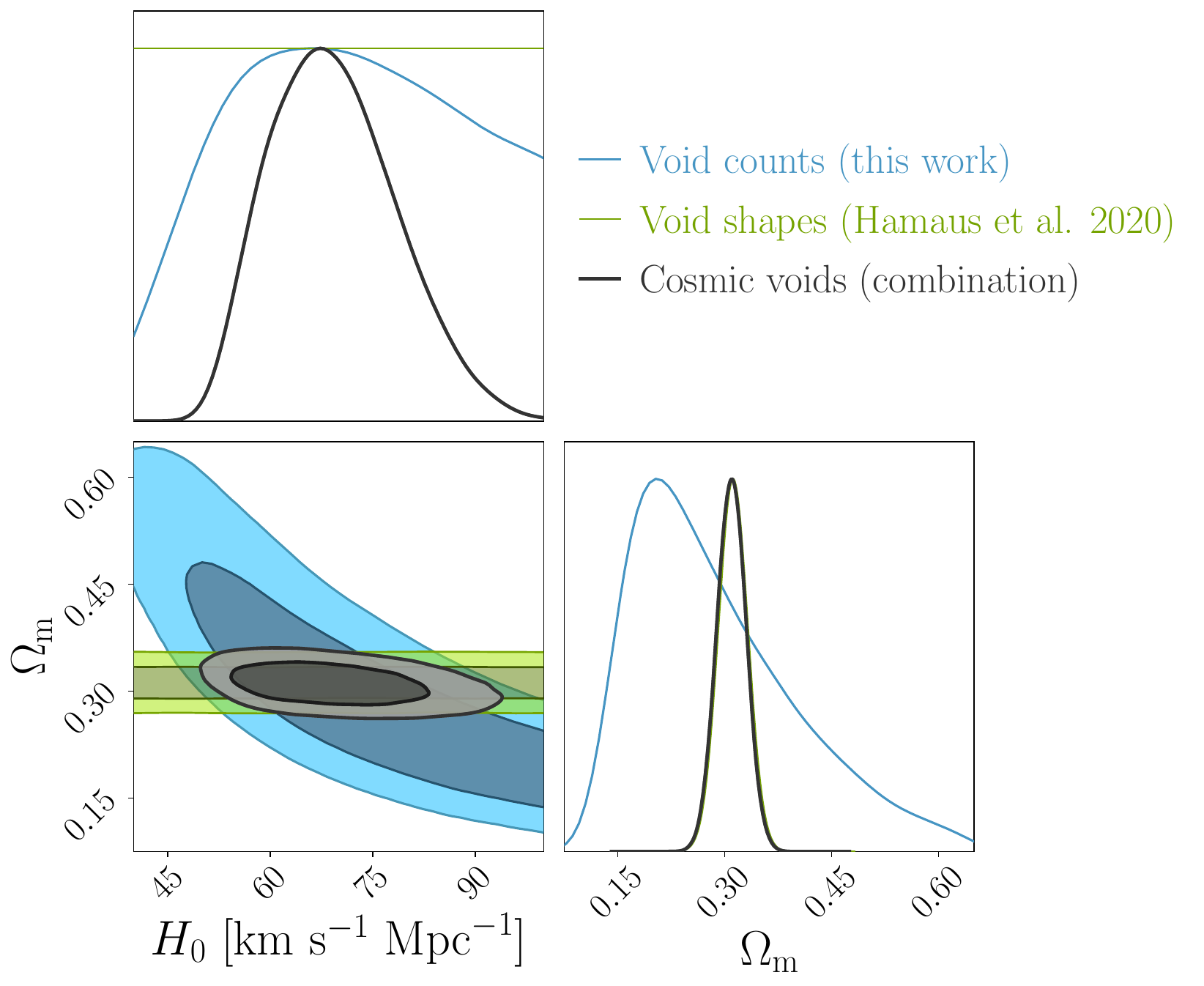}\hspace{0.1cm}
\includegraphics[height=6.5cm]{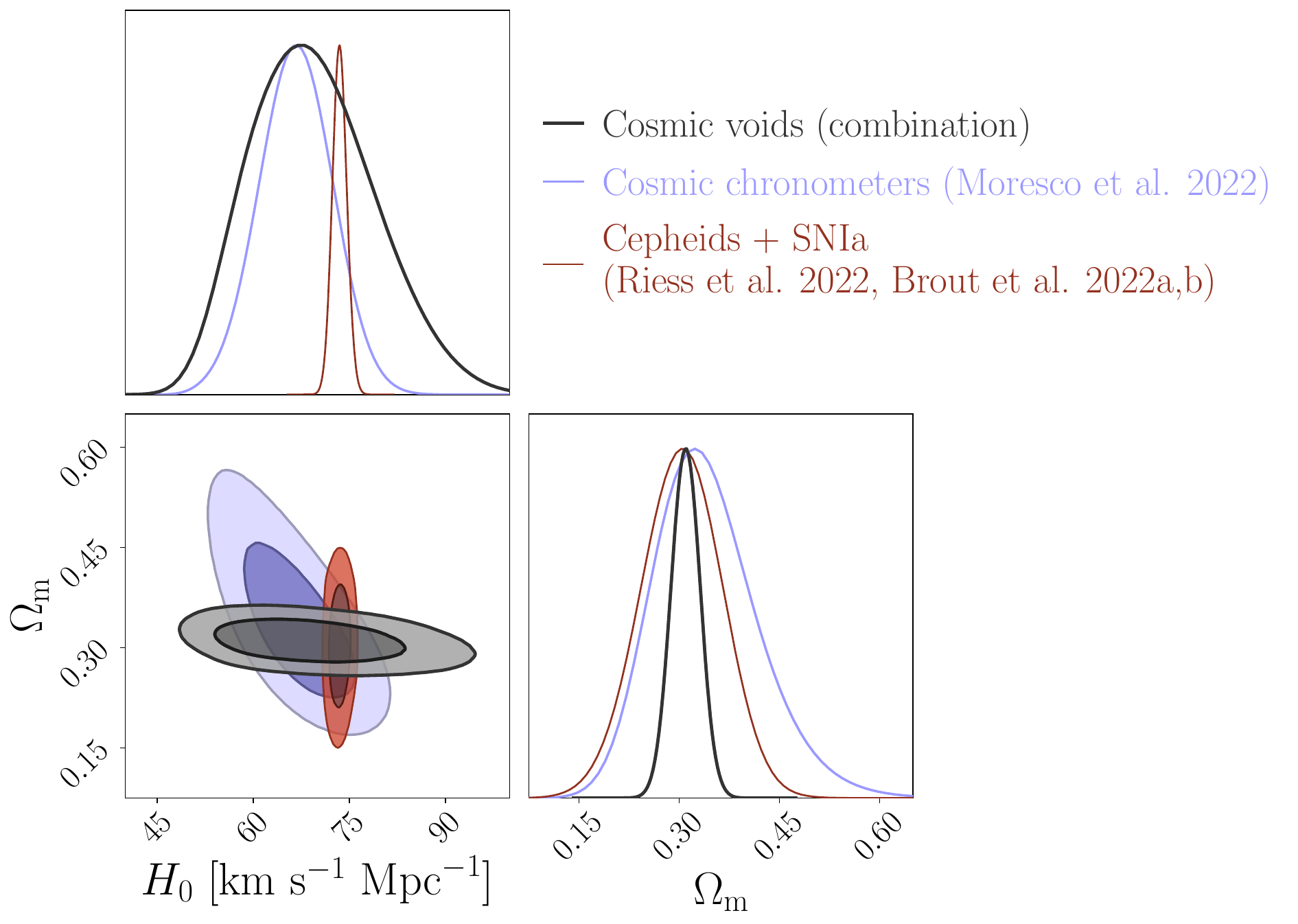}
\caption{Same as \Cref{figS8}, but for the $\Omega_{\rm m}$--$H_0$ parameter plane. In the right panel we now include constraints from cosmic chronometers in violet \citep{Moresco2022} and from the distance ladder in dark red \citep[Cepheids+SNIa,][]{Riess2022,Brout2022a,Brout2022b}. \label{figH0}}
\end{figure*}

The joint analysis of different void statistics we present in this paper is meant to show the full potential of voids as cosmological probes. However, we note that the most valuable gain is expected from the combination of the constraints from the void size function and of those from standard probes, like weak lensing, galaxy clustering and cluster counts \citep{Contarini2022A,Pelliciari}. The expected gain is due to the strong complementarity between the confidence contours (see \Cref{figS8}). A brief explanation for the physical origin of this orthogonality is provided in Appendix D of \cite{Contarini2022BOSS}.

We now move to the analysis of the results for the $\Omega_{\rm m}$--$H_0$ parameter plane.
In \Cref{figH0} we compare these constraints with selected noteworthy results from the literature. In the left panels of this figure, analogously to what we showed in \Cref{figS8}, we report the combination of the cosmological constraints from void counts with those derived from the analysis of void shape distortions \citep[see][]{hamaus2020}. The joint analysis of the two void statistics yields $\Omega_\mathrm{m} = 0.310^{+0.020}_{-0.021}$ and $H_0 = 67.3^{+10.0}_{-9.1} \ \mathrm{km} \ \mathrm{s}^{-1} \ \mathrm{Mpc}^{-1}$.

In the right panels of \Cref{figH0} we compare this result with the constraints obtained by \cite{Moresco2022} via the study of cosmic chronometers extracted from a combination of spectroscopic surveys, and the publicly available\footnote{\url{https://github.com/PantheonPlusSH0ES/DataRelease}} results from the analysis of SN Ia, selected in the Pantheon+ sample \citep{Scolnic2022,Brout2022a,Brout2022b} including the Cepheid host distances and covariance \citep[SH0ES program,][]{Riess2022}. 
Although it shows a large uncertainty with respect to standard probes, our result for $H_0$ represents the very first estimate of the Hubble constant from cosmic voids,  effectively opening the contribution of voids to the landscape of the $H_0$ tension.

\subsection{The void perspective on cosmological tensions}\label{subsec:tensions}

In this section we first compute the value of $S_8$ derived from the modeling of the  BOSS DR12 voids, and then compare it with selected results from the literature. 
From the analysis of void counts alone we derive a value of $S_8 = 0.78^{+0.16}_{-0.15}$, constrained with a precision of roughly $20\%$. 
This relatively large uncertainty is related to the definition of $S_8$; this derived parameter follows the main degeneracy direction of weak lensing and cluster counts measurements, which is unfavorable for perpendicular constraints, such as those obtained in our analysis.
This behavior is naturally mitigated when including the contribution of void shape distortions, as shown in \Cref{subsec:2Dconstraints}. In this case we obtain a value of $S_8 = 0.813^{+0.093}_{-0.068}$, with an accuracy of $9.9\%$.

\begin{figure}[t]
\centering
\includegraphics[width=\columnwidth]{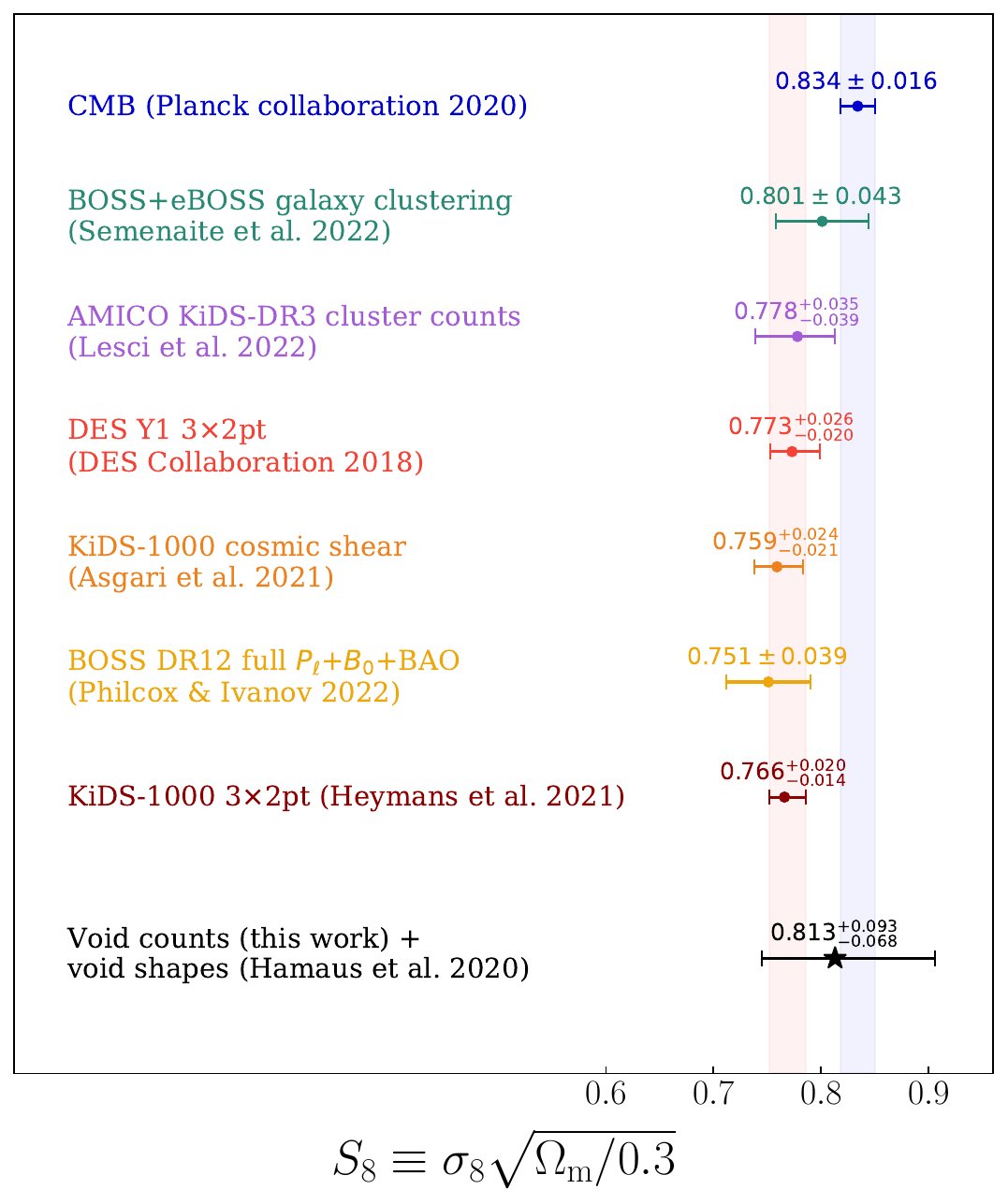}
\caption{Comparison between recent constraints on the parameter $S_8$ from different cosmological probes. The  error bars represent  $68\%$ confidence intervals. From top to bottom: \cite{Planck2018_results}, \cite{Semenaite2022}, \cite{Lesci2022a}, \cite{DES2018_3x2pt}, \cite{Asgari2021}, \cite{Philcox2022}, and \cite{Heymans2021}. The first and the last of these trace two reference bands, in blue and red respectively, highlighting the disagreement between the results at high and low redshift. Our constraint on $S_8$ is  at the bottom (black star with error bars).}
\label{figS8_tensions}
\end{figure}

In \Cref{figS8_tensions} we compare this result with the constraints derived in the following works (from top to bottom): 
\cite{Planck2018_results} from CMB temperature and polarization,
\cite{Semenaite2022} from the full shape of anisotropic clustering measurements in BOSS and eBOSS,
\cite{Lesci2022a} from the mass function of clusters in KiDS DR3, the \cite{DES2018_3x2pt} from the 3x2pt analysis in DES Y1, 
\cite{Asgari2021} from cosmic shear in KiDS-1000,
\cite{Philcox2022} from the large-scale galaxy power spectrum and bispectrum monopole in the BOSS DR12, and
\cite{Heymans2021} from the 3x2pt analysis in KiDS-1000.
Although our constraints cannot statistically exclude any of the $S_8$ estimates reported, we note that the joint analysis of different void statistics provides cosmological constraints competitive with other methods in the literature, and is obtained with an independent methodology.

\begin{figure}[t]
\centering
\includegraphics[width=\columnwidth]{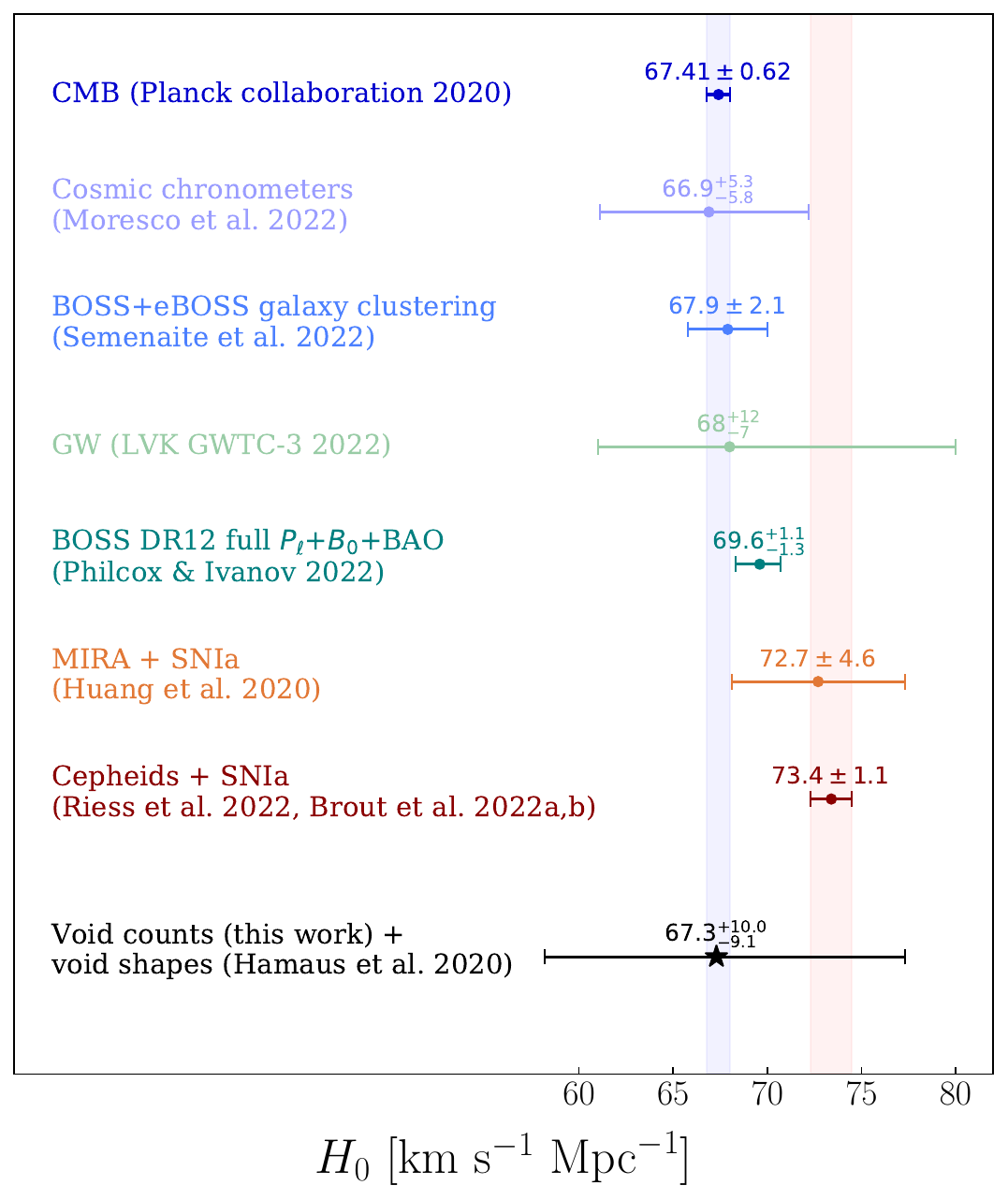}
\caption{As in \Cref{figS8_tensions}, but for the Hubble constant $H_0$. Here  our constraints (in black) are compared with \cite{Planck2018_results}, \cite{Moresco2022}, \cite{Semenaite2022}, \cite{LVK2021}, \cite{Philcox2022}, \cite{Huang2020}, and \cite{Riess2022} and \cite{Brout2022a,Brout2022b}.\label{figH0_tensions}}
\end{figure}

Moving on to the Hubble constant, we consider the constraint resulting from the joint analysis of void counts and shape distortions since only the combination of these two void statistics allows us to break the degeneracy between $\Omega_{\rm m}$ and $H_0$. In \Cref{figH0_tensions} we present a comparison of this result ($H_0 = 67.3^{+10.0}_{-9.1} \ \mathrm{km} \ \mathrm{s}^{-1} \ \mathrm{Mpc}^{-1}$) with recent selected cosmological constraints from the literature. From top to bottom we report the results by the
\cite{Planck2018_results} from the analysis of CMB anisotropies,
\cite{Moresco2022} from the study of cosmic chronometers,
\cite{Semenaite2022} from the anisotropic clustering measurements in BOSS and eBOSS,
\cite{LVK2021} using gravitational-wave sources extracted from the third LIGO-Virgo-KAGRA (LVK) Gravitational-Wave Transient Catalog (GWTC-3),
\cite{Philcox2022} from the full shape analysis of the power spectrum and bispectrum monopole of BOSS DR12 galaxies,
\cite{Huang2020} exploiting the luminosity of a SN Ia calibrated with Hubble Space Telescope Mira variables, and finally \cite{Riess2022} and \cite{Brout2022a,Brout2022b} from the analysis of Pantheon+ SN Ia calibrated with the Cepheids.
Given the large uncertainty, our estimate of $H_0$ from cosmic voids is fully consistent with all the presented cosmological constraints. We expect, however, to considerably improve the voids constraining power in the future by extending the analysis to upcoming survey data and by including the contribution of other void statistics, for example  void lensing and void clustering \citep{Bonici2022, Kreisch2021}.

\section{Conclusions}\label{sec:Conclusions}
In this work we analyzed the size function of voids identified in the BOSS DR12 galaxies \citep{Reid2016}. We computed the void number counts in two redshift bins and  modeled them by means of an extension of the popular Vdn model \citep{jennings2013,contarini2019,Contarini2022BOSS}. The extended model relies on two nuisance parameters, $C_{\rm slope}$ and $C_{\rm offset}$, that we calibrated using 100 realizations of the official BOSS MultiDark PATCHY simulations \citep{BOSSDR12_clustering}. The modeling we considered consistently accounts for redshift-space and geometric distortions and was validated in \cite{Contarini2022BOSS} on the PATCHY mocks.
In a first attempt to exploit the combined power of different void statistics, we computed the constraints resulting from the combined analysis of void counts and shape distortions, the latter coming from \cite{hamaus2020}. The combination procedure we applied in this analysis assumed the two void statistics to be independent.
We estimated the covariance between void counts and the void-galaxy cross-correlation function, revealing that the cross-terms are statistically consistent with zero.

We focused on the $\Omega_{\rm m}$--$\sigma_8$ and $\Omega_{\rm m}$--$H_0$ parameter planes, comparing our results with other selected noteworthy cosmological constraints from the literature, in the light of the modern cosmological tensions. 
First, we presented the results derived from the combination of the two selected void statistics, investigating the ${S_8 \equiv \sigma_8\sqrt{\Omega_{\rm m}/0.3}}$ tension. We previously obtained the marginalized constraints $\Omega_{\rm m} = 0.308^{+0.021}_{-0.018}$ and $\sigma_8 = 0.809^{+0.072}_{-0.068}$, which translate into the derived parameter $S_8 = 0.813^{+0.093}_{-0.068}$. These results are both competitive and compatible with cosmological constraints derived from other probes, such as the void-galaxy cross-correlation function, galaxy cluster number counts, and the 3x2pt statistics, among others.

We then oriented our analysis toward  the $H_0$ tension, deriving from the analysis of different void statistics the marginalized constraints $\Omega_\mathrm{m} = 0.310^{+0.020}_{-0.021}$ and $H_0 = 67.3^{+10.0}_{-9.1} \ \mathrm{km} \ \mathrm{s}^{-1} \ \mathrm{Mpc}^{-1}$. Due to the large uncertainty on the estimate of $H_0$, our constraints are in agreement with various results from the literature, such as those derived from the analysis of CMB anisotropies in the \cite{Planck2018_results} and from SN Ia by \cite{Riess2022} and \cite{Brout2022a,Brout2022b}.

At this stage the presented constraints on the $S_8$ and $H_0$ parameters do not allow us to take a decisive position in the context of the current cosmological tensions. Nevertheless, this work shows for the first time the relevance of cosmic voids in the context of modern tensions in cosmology, providing an independent probe. We plan to extend our analysis to the data of upcoming wide-field surveys such as \textit{Euclid} \citep{Laureijs2011, Amendola2018}, the Dark Energy Spectroscopic Instrument \citep{DESI2016}, the Prime Focus Spectrograph \citep[PFS,][]{PFS_2016},
the Nancy Grace Roman Space Telescope \citep[NGRST,][]{spergel_2015_WFIRST}, the Spectro-Photometer for the History of the Universe and Ices Explorer \citep[SPHEREx,][]{SPHEREx_2018},
and the Vera C. Rubin Observatory \citep[][]{LSST_ivezic2019}.

\vspace{2mm}

\begin{acknowledgements}
SC thanks Michele Moresco, Alfonso Veropalumbo, Giorgio Lesci and Nicola Borghi for the contribution and suggestions they provided. AP is grateful to Scott Dodelson for useful discussions. The authors thank Steffen Hagstotz, Barbara Sartoris, Nico Schuster, Giovanni Verza and Ben Wandelt for useful conversations. We acknowledge the grant ASI n.2018-23-HH.0. SC, FM, LM and MB acknowledge the use of computational resources from the parallel computing cluster of the Open Physics Hub (\url{https://site.unibo.it/openphysicshub/en}) at the Physics and Astronomy Department in Bologna. AP is supported by NASA ROSES grant 12-EUCLID12-0004, and NASA grant 15-WFIRST15-0008 to the Nancy Grace Roman Space Telescope Science Investigation Team ``Cosmology with the High Latitude Survey''. AP acknowledges support from the Simons Foundation to the Center for Computational Astrophysics at the Flatiron Institute. NH is supported by the Excellence Cluster ORIGINS, which is funded by the Deutsche Forschungsgemeinschaft (DFG, German Research Foundation) under Germany's Excellence Strategy -- EXC-2094 -- 390783311. LM acknowledges support from PRIN MIUR 2017 WSCC32 ``Zooming into dark matter and proto-galaxies with massive lensing clusters''. We acknowledge  use  of  the Python libraries \texttt{NumPy} \citep{numpy}, \texttt{Matplotlib} \citep{Matplotlib} and \texttt{ChainConsumer} \citep{ChainConsumer}. 
\end{acknowledgements}

\bibliographystyle{aa}
\bibliography{bibliography}

\end{document}